**Adding fuel to human capital: Exploring the educational effects of cooking fuel choice from rural India**

Shreya Biswas
Assistant Professor, Department of Economics
Birla Institute of Technology Pilani, Hyderabad Campus
(Email: shreya@hyderabad.bits-pilani.ac.in)

Upasak Das
Presidential Fellow in Economics of Poverty Reduction
Global Development Institute, University of Manchester
And
Research Affiliate
Centre for Social Norms and Behavioral Dynamics
University of Pennsylvania
(Email: upasak.das@manchester.ac.uk)

**Please do not cite without permission**

**June 3, 2021**

**Adding fuel to human capital: Exploring the educational effects of cooking fuel choice from rural India**

***Abstract***

*The study examines the effect of cooking fuel choice on educational outcomes of adolescent children in rural India. Using multiple large-scale nationally representative datasets, we observe household solid fuel usage to adversely impact school attendance, years of schooling and age-appropriate grade progression among children. This inference is robust to alternative ways of measuring educational outcomes, other datasets, specifications and estimation techniques. Importantly, the effect is found to be more pronounced for females in comparison to the males highlighting the gendered nature of the impact. On exploring possible pathways, we find that the direct time substitution on account of solid fuel collection and preparation can explain the detrimental educational outcomes that include learning outcomes as well, even though we are unable to reject the health channel. In the light of the micro and macro level vulnerabilities posed by the COVID-19 outbreak, the paper recommends interventions that have the potential to fasten the household energy transition towards clean fuel in the post-covid world.*

***Keywords:*** *Solid fuel, time use, rural India, education, gender*

***JEL Code:*** *I18; I31; I21*

## 1. Introduction

A substantial proportion of households especially across the Global South depends on solid fuel as a primary source of cooking fuel. The World Health Organization (WHO) in 2018 estimates about three million people not having access to clean fuel for household usage emphasizing the need for a faster energy transition to clean fuel. Literature has indicated attributes like wealth, income, fuel price, education, gender and preferences to be among the major determinants of household fuel choice (Behera et al., 2014; Muller and Yan, 2018; Rahut et al., 2020). Despite major gains in the adoption of cleaner fuel across the developing countries, the onset of the COVID-19 pandemic is further likely to push households who had transitioned to clean fuel back to the usage of non-cleaner options because of reasons that include significant income reduction, job-loss and cut in subsidy among others (Zhang and Li, 2021).[1] The reliance on solid fuel for cooking is related to negative externalities on environment and health across developing countries. Accordingly, the Sustainable Development Goals (SDG) have recognized its importance and added "ensure universal access to affordable, reliable and modern energy services" as one of its goals to be achieved by the year, 2030.

Literature has indicated significant detrimental effects of using solid cooking fuel on health outcomes. Studies have documented lower birth weight, increased likelihood of developing respiratory infection and higher risk of mortality among children who are exposed to household air pollution caused through solid fuel use (Edwards and Langpap, 2012; Epstein et al. 2013; Naz et al. 2016). Other studies have shown an increased likelihood of health issues that include breathing problems, ophthalmic issues and blood pressure among others, which are associated with higher household air pollution (Jagger and Shively, 2014; Neupane et al. 2015; Arku et al. 2018; James et al. 2020). In fact, the ability to cope up with daily activities among the elderly is found to reduce significantly in these households (Liu et al. 2020). This paper adds and also complements the existing evidence on the welfare effects of cooking fuel choice on an important albeit less explored dimension that is the human capital investments through education among adolescent children from rural India. Additionally, we also explore the extent to which these effects are gendered and the associated mechanism.

[1] https://www.eco-business.com/news/delhis-poor-return-to-dirty-fuel-in-covid-19-lockdown/ (accessed on May 12, 2021)

Cooking fuel choice and education may be linked in multiple ways and also depends on the context. In India, the majority of the households not using cleaner fuel depend on firewood or other natural resources that include cow dung or crop residues for the purpose of cooking. Accordingly, a significant amount of time gets allocated for the collection of these resources. Studies have indicated that school-going children can potentially substitute their school time with activities involving collection of these resources which in turn can affect their school performance as well (Ndiritu and Nyangena, 2011; Levison et al. 2017). As resource gets scarcer, the demand for these "environmental chores" increases and as a result parents may trade-off investment in human capital with higher labor requirements from the children. Extant literature has indicated a strong association of resource collection on school attendance and years of education among children in different contexts (Nankhuni and Findeis, 2004; Ndiritu and Nyangena, 2011; Levison et al. 2017). Importantly if these activities are placed alongside domestic responsibilities, the educational effects are likely to be gendered. This is because of a disproportionately higher time spent on collection of fuel-wood by females which is driven by the traditional division of labor and expected household roles (Choudhuri and Desai, 2021; Nankhuni and Findeis, 2004). In addition, it is also possible that the adverse health effects of solid fuel may affect schooling and academic performance of children. In this paper, while we find evidence of time substitution effects owing to firewood and other natural resource collection, we are unable to reject adverse health effects also driving the deterioration in educational investments.

Studying the implication of cooking fuel choice on educational outcomes in rural India is particularly important for multiple reasons. Firstly, India has among the largest population who are dependent on solid fuel for cooking purposes, many of whom are located in rural areas. Data from the Census of India, 2011 shows about 62.5 of rural households are dependent on firewood for cooking, followed by more 12 percent who depend on crop residue. This has contributed to a higher concentration of household pollutants and accordingly has been recognized as among the most important risk factors responsible for mortality and morbidity in India (WHO, 2018). Field experiment suggests that improvement in economic wellbeing is related to an increase in energy consumption of poor households in accordance to the energy ladder theory but does not necessarily shift the preferences towards clean cooking fuel (Hanna and Oliva, 2015). To accelerate the transition from dirty to clean fuel, the Pradhan Mantri Ujjwala Yojana (PMUY) was implemented by the Government of India in 2016 that promised subsidized Liquefied Petroleum Gas (LPG) to poor households has been successful initially in energy transition. However studies have found evidence in favor of fuel stacking and raised questions with respect to the extent and regular usage of LPG with many households resorting to traditional fuel because of easy access, lower price, difficulties in refilling cylinders and lack of awareness (Kar et al. 2019, Swain and Mishra, 2020). Second, in terms of educational outcomes among adolescent children, while significant progress has been made in the attainment of primary education, completion of lower and higher secondary education remains elusive. In 2016-17, the gross enrolment ratio at lower secondary level was found to be around 78.5% while it was about 52% at the senior secondary level (Tilak, 2020). The same study also documents high dropout among children in secondary and higher secondary grades, which is more pronounced for females. Additionally, despite completion of eight years of schooling, a significant proportion is found to be unable to read a basic text and perform simple 3 by 1 division (World Bank, 2018; Government of India, 2019).

A number of studies have looked into the effect of fuel poverty on children wellbeing and have put forward the effects on academic performance as a channel (Churchill et al. 2020; Zhang et al. 2021). In fact, a recent study by Choudhuri and Desai (2021) has found adverse educational outcomes among rural Indian children from households that depend on free collection of water and cooking fuel. However, the study focuses particularly on young children in the age cohort 8 to 14 years and looks at cognitive abilities for those from 8 to 11 years. Our study complements this study, by examining the effect of fuel on choice on adolescent children in the age cohort 12 to 18 years rather than focusing on children studying in the primary and upper-primary level. As mentioned, because higher school dropout among children in India happens at secondary and higher secondary level, we argue the implications of cooking fuel choice can potentially be more pronounced for children in the 12-18 age group relevant for this educational level. More importantly, if the effect of education on job opportunities is non-linear in nature, educating children till primary level may not fetch superior labor market outcomes as against completing secondary or higher secondary education. With respect to time allocation for firewood and other natural resource collection, primary school-going children would be dependent on elder family members who may substitute their time away from childcare, thus indirectly leading to deteriorated educational outcomes. In contrast, adolescent children are more likely to allocate this time themselves which can directly have educational consequences. Interestingly, Choudhuri and Desai (2021) also document lower mathematical scores among boys because of higher psychosocial fragility due to lack of motherly supervision. This childhood disadvantage may then result in further worsening of outcomes for adolescent males. However, because women typically bear a greater burden of household chores that include fuel resource collection, it may result in disproportionately higher crowding out of schooling for females. In terms of policy implications, our paper assumes significance as studies have found a direct causal link of lower educational outcomes among adolescent females on labor market opportunities apart from improving relative bargaining power, which can lead to an increase in age at marriage, higher autonomy in fertility decisions and agency among others (Rindfuss and St. John, 1983; Field and Ambrus, 2008; Jensen, 2010; Abalos, 2014; Asadullah and Wahhaj, 2019).

To assess the implication of cooking fuel choice and time allocation of free collection of firewood and other natural resources on educational outcomes, this paper uses multiple representative datasets which are relevant for studying these issues. Firstly, to estimate the effects on school attendance and years of education, we use the fourth round of the National Family Health Survey (NFHS-4) conducted by the Ministry of Health and Family Welfare, Government of India in 2015-16 which collects household data on cooking fuel use and education along with that on a host of socio-economic characteristics. To address the concerns revolving around unobservable confounders because of selection bias, we utilize exogenous variation in local forest cover in the lagged period that can be assumed to be highly correlated with cooking fuel choice but not directly related to educational outcomes. For this purpose, we use data from the Socioeconomic High-resolution Rural-Urban Geographic Dataset (SHRUG) that records yearly forest cover using vegetation continuous field, a MODIS product. For examining the implications of time allocation for resource collection linked with households using solid fuel for cooking, we use the Time-Use Survey (TUS) data collected in 2019. Further, we gauge the potential impact on learning outcomes using the Annual Status of Education Report (ASER) survey data conducted in 2016 among children.

We find significantly adverse educational effects of solid fuel use on adolescent children measured in terms of their likelihood of attending school, years of education and age-appropriate grade attainment. The results are found to be robust despite accounting for the potential unobservable confounders that we ensure through usage of exogenous variation in past forest cover as instruments. Importantly, our causal inference stand even when we change the instrument, the age cohorts considered or definition of solid fuel along with alternate estimations that allow for the instruments to be “plausibly exogenous”. Additionally, we observe that these educational effects are likely to be more adverse for females. On exploring the potential channels, we find time spent on schooling and home-work to be significantly lesser among children who allocate more time for the collection of firewood. Notably, the substitution effects are disproportionately higher for females implying that household cooking fuel choice is likely to affect them more not only with respect to time spent in school but also that dedicated for doing homework. If this is case, it is possible that the learning outcomes would also be adverse. This is exactly what is observed as we find children from districts with a higher proportion of households using solid fuel are found to score worse and these ill-effects are again disproportionately higher for females. Importantly, among the children considered, we observe the female disadvantage to less prominent among younger children while it is found to diverge for older children.

The paper has multiple contributions. Firstly, to our knowledge, this is among the first few papers to present robust empirical evidence of how cooking fuel choice is linked to educational outcomes. Despite previous studies documenting the impact of fuel poverty on education and children wellbeing (Churchill et al. 2020; Choudhuri and Desai, 2021; Zhang et al. 2021), this paper estimates the educational effects on adolescent children and studies how these effects are gendered with age. Additionally, it provides evidence of direct time substitution effects of time allocation for fuel resource collection. Secondly, the paper goes beyond the common determinants of educational outcomes that include individual, household and school level factors among others and shows the potential implications that may arise because of cooking fuel choice. Third, the paper complements the expanding literature that outlines direct and indirect effects of solid fuel usage. Apart from the adverse effects of solid fuel choice on the environment and health, it argues that dirty fuels may lead to detrimental human capital outcomes for children, which in turn may perpetuate intergenerational poverty. Additionally, because the educational effects are found to be disproportionately high for females, one can also link longer-term adverse effects that include lower age at marriage, female empowerment and labor participation among others with solid fuel usage. Accordingly, the paper underscores the importance of implementing robust policies that incentivize households to switch from solid fuels to cleaner options.

The remainder of the paper is structured as follows. Sections 2 and 3 present the data and the empirical strategy, respectively. Section 4 presents the main results and associated robustness tests. Section 5 explores the possible explanations for the observed results.

Finally, section 6 discusses the importance of the findings in the light of the ongoing coronavirus pandemic and section 7 summarizes the finding along with the policy implications.

## 2. Data

We employ the NFHS-4 conducted during January 2015 and December 2016 by the Ministry of Health and Family Welfare, Government of India to analyze the effect of fuel type on educational effects. The survey gathered information on 601,509 households across 640 districts and is one of the largest household surveys in India that is representative at the national, state and district level. The survey provides information related to cooking fuel type, educational attainment of individuals residing in the household, school attendance, age along with other characteristics like economic status, religious and caste affiliation of the households.

The forest cover data is obtained from SHRUG India platform, which is an open dataset providing high resolution data on forest cover, night lights, employment and political outcomes among others at village/ town level. SHRUG provides total forest cover data (Asher et al., 2021) from 2000 to 2019 aggregated from Vegetation Continuous Fields at 250m resolution using georeferenced location polygons generated from machine learning models.[2] The SHRUG database has been used by several studies ranging from the effect of road construction to influence of political leaders on economic outcomes (Asher and Novosad, 2017; Lehne et al., 2018; Adhukia et al., 2020).

Further, we use the TUS conducted in 2019 by the National Sample Survey Organization of India. The survey collects information about the time allocation of different activities undertaken by the members of the sampled households on the day prior to the survey. It gathered data from 518,751 individuals residing in 138,805 households. This survey is representative at the national as well as at the state level and can be useful in understand time use and allocation patterns across individuals. This survey data in our paper is primarily used to examine the time substitution effects from schooling to that allocated for firewood and other natural resource collection in households that use solid fuel for cooking purposes.

---

[2] This data is retrieved from Dimiceli, C., Carroll, M., Sohlberg, R., Kim., D., Kelly, M., & Townshend, J. (2015). MOD44B MODIS/Terra Vegetation Continuous Fields Yearly L3 Global 250 m SIN Grid V006 [Data Set]. *NASA EOSDIS Land Process*.

To examine the implications of cooking fuel choice on learning outcomes, we employ household survey data from the ASER form the year 2016.3 We particularly use the 2016 household survey data as we use it along with NFHS 4 data, which was also conducted in 2015-16. The survey is conducted to analyze the enrollment status as well as the basic learning levels among children from rural India and is representative at the district level. The survey covered about 17,473 villages across 589 rural districts and collected data from 350,232 households. Using well tested rigorous tools, the ASER survey collects information on basic mathematics and reading proficiency levels from all children in the age group 5 and 16 years, who are residing in the sampled households.4 These outcomes have been used extensively by other studies (Chakraborty and Jayaraman, 2019; Lahoti and Sahoo, 2020). The description of the variables used in the study is provided in Table 1.

[Insert Table 1 here]

## 3. Estimation strategy

We estimate the following equation to explore the educational effects of non-clean cooking fuel usage:

$$Y_{is} = \alpha + \beta SF_{is} + \rho X_{is} + \mu_s + u_{is} \qquad (1)$$

Here $Y_{is}$ is the educational outcome for the child, $i$ residing in state, $s$. These outcomes include years of education and age-appropriate grade completion among others as defined in Table 1. $SF_{is}$ is a binary variable that takes the value of 1 if the corresponding household of child, $i$ uses solid fuel for cooking purpose and 0 otherwise. $X_{is}$ is the set of individual and household level characteristics that can be hypothesized to determine educational outcomes as given in Table 1. $\mu_s$ is the state level fixed effect that accounts for the heterogeneities at the state level. This variable controls for attributes like differential state policies on education, female autonomy or rural development among others. $u_{is}$ is the error term in the model.

As one may argue, direct Ordinary Least Squares (OLS) estimation may lead to biased estimates because of unobserved factors confounding the casual estimates. It is possible that households located in areas with poorer administrative efficiency would have lower schooling owing to a greater distance to school and also have lower access to clean fuel. Further, households with higher bargaining power among women may be more likely to adopt clean fuel and simultaneously invest in higher human capital for children. To address these endogeneity concerns, we use exogenous variation in average forest cover at the district level that can be hypothesized to be correlated with non-clean fuel usage but not related to children educational outcomes through channels other than solid fuel usage. Higher forest cover in the vicinity ensures better and easy access to firewood and hence may discourage households to adopt cleaner fuel (Bhat and Sachan, 2004; Tembo et al., 2015). Accordingly, we argue households situated in districts with higher forest cover are more likely to use firewood for cooking purposes. Of note is the fact that firewood is the most popular choice of cooking fuel among households using the non-cleaner options. The NFHS conducted in 2015-16 indicates 78% of rural households who use solid fuel are dependent on firewood.[5] This indicates how overwhelmingly firewood dominates the cooking fuel space among households not using cleaner options and hence average forest cover is likely to be strongly linked with its usage.

[3] For more information, refer img.asercentre.org/docs/Publications/ASER Reports/ASER 2016/aser_2016.pdf (accessed on May 25, 2021)
[4] These tools can be accessed from http://www.asercentre.org/p/141.html (accessed on May 24, 2021)

However, educational outcomes are unlikely to be related to forest cover when other possible channels are accounted for. One can still argue that unobservables may still drive households in self-selecting themselves to reside in locations with higher forest cover and these unobservables may also influence educational outcomes. However, studies have indicated that spatial mobility is low in India because of which residential relocation should not be a major cause of concern in our case (Munshi and Rosenzweig 2016; Rowchowdhury 2019). Nevertheless we still allow for the possibility of instruments not meeting this exclusion criterion through the estimation strategy developed by Conley et al. (2012) which assumes the instrument to be "plausibly" but not fully exogenous. This has been detailed out in section 4.3.2. In addition, we also use Propensity Score Matching (PSM) with Rosenbaum bounds (section 4.3.3) to ensure that the estimates are not confounded by unobservables and hence unbiased.

## 4. Results

### *4.1 Descriptive statistics*

Table 2 presents the summary statistics for the selected variables for the full sample and for solid and non-solid fuel users. Notably, the educational outcomes for solid fuel users are poorer than non-solid fuel users and this difference is statistically significant at 1% level of significance. Further, the solid fuel users have a larger household size, lesser number of adult household members with completed primary schooling, less likely to have access to piped drinking water or toilet facility at home compared to non-solid fuel users. As one would expect, these households are also less likely to own mobile and are poorer as indicated by the share of households owning a Below Poverty Line (BPL) card.

---

[5] This is unweighted figure from the survey.

[Insert Table 2 here]

Figure 1 presents the distribution of years of education for the 12 to 18 years old children based on the type of cooking fuel. Visual inspection suggests that the average years of schooling is higher for children from non-solid fuel user households in comparison to those belonging to solid-fuel user households. Given that other confounders may drive this difference between the groups, we explore this further using the regression framework.

[Insert Figure 1 here]

*4.2 Main results*

Table 3 presents the baseline association between solid fuel and educational outcomes. As mentioned, we use three main outcomes: school attendance, years of schooling and age appropriate grade progression and regression estimations of these three indicators are given in columns 1 to 3 respectively. The findings reveal usage of solid fuel is related to lower likelihood of attending school (Column 1), lesser years of schooling on average (column 2) and also lower average grade progression (column 3). Therefore, the preliminary analysis suggests a negative and statistically significant relationship between solid fuel use and educational outcomes of children in India.

[Insert Table 3 here]

As acknowledged earlier, the solid fuel variable can be endogenous because of self-selection bias wherein households who use solid fuel can also have lower human capital investment for a variety of reasons other than that related to cooking fuel. Accordingly, endogeneity corrected 2SLS estimation are used to obtain the causal inference. Columns 4-6 presents the corresponding results for the three outcome variables using average forest cover six years before the survey year at the district level as an IV. Specifically, we find that using solid fuel leads to a reduction in the likelihood of attending school by 18.2 percentage points for the rural children in the age-cohort 12 to 18 years (column 4).[6] Further, solid fuel is also found to be associated with 1.01 standard deviations lower years of schooling for these adolescent children on average, which is also found to be statistically significant at 1% level (column 5). Complementary to these findings, we also observe a 0.31 slower grade progression of children compared to non-solid fuel users (Column 6), ceteris paribus. These results indicate a definite discernible educational loss among rural adolescent children because of cooking fuel choice. Importantly, the first stage regression suggests that there is a positive and significant relationship between average forest cover and solid fuel use. The first stage F-statistic is much higher than the commonly accepted threshold of 10 in all the specifications, suggesting that the IV is not a weak one. Because we find the marginal effects from the OLS regression to be much smaller than those from the 2SLS regression, naïve OLS estimates without accounting for the endogeneity bias could have possibly underestimated the adverse educational effects of solid fuel use.

The results for the covariates are in line with the existing literature and holds with respect to the context of rural India (Drèze and Kingdon, 2001; Tilak, 2020).[7] Girls are found to be less likely to attend school and have lower years of schooling. As one would expect, household educational outcome is positively related to child's educational outcomes. A larger family size is associated significantly with inferior educational attainment which is also true for the Muslim children. The children from upper caste households and households possessing mobile phones have significantly better educational outcomes on average.

### *4.3 Robustness checks*

We perform a battery of robustness checks to ensure that our results capture the effect of solid fuel use on educational attainment and are not confounded by other factors that we are unable to capture.

#### 4.3.1 Time use survey

The nature of the TUS, the details of which has been explained in section 2 allows us to examine the educational effects of solid fuel usage using a different dataset, which is representative. Further, because the TUS has been conducted in 2019 as against the NFHS-4 which was administered in 2015-16, it enables us to revisit using more recently gathered household data. Additionally, apart from the educational attainment indicator, the TUS also collects information on the amount of time allocated for formal education. We use these indicators (in standardized form) as outcome variables and estimate the OLS along with the 2SLS regression with average forest cover six years back (year: 2013) as the IV. The control scheme is kept similar to that used for earlier regression with the only difference being the introduction of household monthly consumption expenditure instead of possession of BPL card. Information on BPL card possession is not collected in the TUS dataset and hence we add consumption expenditure as the latter is a well-accepted indicator of household economic conditions (Deaton, 2005). Findings from the regression are presented in appendix table A3. We observe robust but adverse influence of household solid fuel usage not only on educational attainment but also time spent in formal education.

[6] The reported results are based on LPM models in both first and second stages. The instrumental variable probit estimates are presented in table A1 of the appendix.

[7] The regression results for all the covariates are given in appendix A2

#### 4.3.2 Plausibly exogenous

One may argue that the IV we use does not satisfy the strict exogeneity condition and the 2SLS estimation results are biased due to this violation of this assumption. In our context, for example, it is possible that districts with higher past average forest may systematically experience rapid deforestation and industrialization which may potentially improve employment opportunities. This in turn may increase the opportunity cost of schooling among adolescent children. To avert these possibilities confounding our estimates, Conley et al. (2012) suggest ways to draw causal inference when the instrument is only ‘plausibly exogenous’. In this framework, we consider the equation below:

$$y_i = \beta SF_i + \gamma z_i + \emptyset X_i + \varepsilon_i \qquad (2)$$

Where $y_i$ are the three educational outcome variables, $SF_i$ refers to the endogenous solid fuel variable and $z_i$ is the instrument past average forest cover and $X$ is the vector of controls. The main results discussed in section 4.2 assume that $\gamma$=0. If the parameter $\gamma$ is close to zero, but not necessarily zero, the instrument can still be ‘plausibly exogenous’ and it is possible to consistently estimate the effect of the endogenous variable on the outcome variable for certain values of $\gamma$. Extant literature has used this in different contexts (McArthur et al., 2017; Azar et al., 2021; Das, 2021). Following Azar et al., (2021), we estimate the reduced form equation of educational outcomes on the instrument and other control variables, but excluding the endogenous solid fuel variable and obtain the lower bound for $\hat{\gamma}$ (Table 4). Using values of $\gamma$ in the range of $\hat{\gamma}$ to zero, we obtain the bounds for the effect of solid fuel in the second-stage. Further, we also report the maximum value of $\gamma$ ($\gamma_{max}$) for which the estimated bound for $\beta$ is strictly less than zero. Our estimated $\gamma_{max}$ values indicate that solid fuel usage will have adverse effects on the educational outcomes, even if the direct effect of the instrument is up to 36%--75% of the reduced form effect. The results we obtain are robust to a greater degree of the instrument exogeneity assumption relaxation for standardized years of schooling and grade progression variables and to a lesser extent to school attendance variable. Accordingly, even if the instrument is not fully exogenous i.e. $\gamma$ is non-zero, it is unlikely that the endogenous solid fuel variable will not even capture 25% for years of education and grade progression effects, respectively. For school attendance, the endogenous should capture at least 64% of the combined effect of the instrument and the endogenous variable. The result of the plausibly exogenous IV technique reinforces that the IV causal estimates are robust to allowance of a substantial relaxation of the exclusion restriction in our regression model.

[Insert Table 4 here]

4.3.2 Propensity score matching (PSM)

As an additional robustness check, we employ PSM to compare the difference in the educational outcomes of children between solid and non-solid fuel users. PSM can be employed to address endogeneity and draw causal inferences using observational data in a non-experimental set up (Dehejia and Wahba, 2002). Importantly existing literature that examines the cooking fuel/ energy poverty effects on welfare outcomes have extensively used PSM either as the main econometric method to decipher the effects or as a part of robustness exercise (Churchill et al., 2020; Liu et al., 2020). The treatment here is solid fuel use the matching is performed using a set of observed covariates. We consider the nearest neighbor matching method as proposed by Rosenbaum and Rubin (1985) with a caliper of 0.01 and obtain the average treatment effect on the treated. Table 5 suggests all our outcome indicators including the likelihood of attending school, average educational attainment and age appropriate grade progression among children from solid fuel user households are lower than that from the non-solid fuel users at 1% level of significance post-matching. Of note is the fact that the results of the PSM approach corroborate our findings of the 2SLS IV estimates that we presented.

One of the key drawbacks of the PSM method is it accounts for endogeneity only due to observed covariates while omitted variables that affect both solid fuel adoption and educational attainment may still confound the estimates. Accordingly, we report the Rosenbaum bounds that indicate that the PSM results are robust in the presence of hidden bias up to a threshold. Rosembaum (2002) developed an approach to assess the effect of bias on test statistic if the assignment to treatment is not random owing to unobservable factors using a non-parametric Wilcoxon signed rank test. Column 4 indicates that the Rosenbaum bounds in the range of 1.28-1.39 for the various outcome variables.[8] Specifically, it suggests that hidden bias in the range of 28-39% will still yield a significant difference in educational outcomes among solid and non-solid fuel users.

[Insert Table 5 here]

4.3.4. Additional sensitivity analysis

Next, instead of district level forest cover six years before the survey, we consider the variable five years back as an instrument. This alternate instrument yields qualitatively similar results in the 2SLS regression wherein we find that household traditional fuel usage adversely affects educational outcomes for adolescent children at 1% level of significance (columns 1-3, appendix tableA4). The results are also robust if we consider 11 to 18 as adolescent children, respectively and re-estimate equation 1 (columns 4-6). Finally, we reclassify kerosene as a source of dirty fuel and re-estimate the 2SLS regression. This is because there are studies that indicate that kerosene is less polluting than other solid fuels like coal or firewood, but is not as clean as modern fuels like LPG or electricity (Smith et al., 2000; Thoday et al., 2018). The findings from all these regressions indicate robust and significant but negative educational effects on children from household using dirty fuel (columns 7-9).

## 5 Mechanisms and further analysis

### *5.1 Is there a gendered effect?*

Our regression results till now establish an educational loss for an average adolescent child that is linked with the household cooking fuel choice. However, are these effects similar for boys and girls? Who, among the children are more likely to be disproportionately affected due to solid fuel usage? We explore these questions by examining the impact separately for the females and males in our sample. Figure 2 presents the 2SLS IV regression results for children of the same age cohort (12 to 18 years). The marginal effects associated with the solid fuel variable is found to be negative and significant at a 5% level for both females and males in all the specifications; however, the results are most striking for years of schooling and grade progression variables with a discernible disadvantage for the females. These systematic higher adverse effects observed for females in comparison to the males indicate a definite gendered pattern in fuel type and education relationship among children in India.

[8] Liu et al. (2020) find Rosenbaum bounds in a similar range.

[Insert Figure 2 here]

### *5.2 Time substitution*

If female education suffers disproportionately in comparison to the males because of household cooking fuel choice, one key channel due to which this can potentially happen is through time substitution of that allocated to educational activities. Literature has pointed out that activities including collection of firewood and processing of dung cakes for cooking and other household use are common in rural India (Hirway and Jose, 2011; Choudhuri and Desai, 2021). Since women take responsibility of these activities largely, it is possible that the allocation of time on educational activities gets substituted more for females systematically affecting their schooling outcomes relative to the males. This has been discussed extensively in the developing country context including India (Nankhuni and Findeis, 2004; Ndiritu and Nyangena, 2011).

The TUS data allows us to examine these effects for the children we consider in the age cohort 12 to 18 years. Here we consider total time spent in school and that on doing homework as the outcome variables with total time allocated for firewood and other natural resource collection as the main variable of interest. To ensure we can get close to obtaining causal estimates of time allocation to fuel resource collection on the mentioned outcomes, we use three specifications to control for all possible confounders. In the first specification, we include all the set of control variables as indicated earlier. In the second one, we include time allocated for collection of fuel resource by others in the household. This would account for indirect effects, if any on children schooling that may emanate because of time substitution away from child care due to firewood and other cooking resource collection by others members of the household (Choudhuri and Desai, 2021). As one may note, because of this, it is possible that the adolescent children then have to substitute time away from school to domestic chores that include caring for elderly and the kids. Consequently, in the third specification, we control for time allocation for caring activities by the corresponding child. Please note that only rural children in the age cohort 12 to 18 years from households using solid fuel are considered. This is because free collection of firewood and other natural resources would only be relevant for households using solid fuel and not for those using the cleaner options.

We first analyze the marginal effects from regressions using male sample following it up with the female sample, the results of which are given in appendix figures A5 and A6, respectively. We observe a significant reduction in school and homework time as the time allocated for fuel collection increases for males as well as females. Nevertheless, the effect size indicates that the females are likely to be more affected with the time substitution because of the collection of fuel. We check this through a regression of school and homework time on an interaction variable of the female dummy with the solid fuel dummy (Figure 3). The negative and statistically significant association between the two indicates that female children are likely to be systematically more affected on average than male children. Importantly, this relationship remains intact in all specifications and additionally, we do not observe any discernible change in the effect size as well.

[Insert Figure 3 here]

*5.3 What about performance?*

If school and homework time get substituted because of higher time allocation for collection of firewood and other natural resources, learning outcomes or cognitive abilities among children are likely to suffer as well. While NFHS-4 or TUS do not directly collect information on learning outcomes, ASER dataset collects relevant data on this as mentioned earlier. Accordingly, we use this dataset to draw inference on the potential association of cooking fuel choice on children learning outcomes. The reading skill assessment that ASER administers has the following four levels which are ordinal in nature: recognition of letters, reading of words, reading a short paragraph (a grade 1 level text), and reading a short story (a grade 2 level text). The arithmetic skill assessment also has four ordinal levels: recognition of single-digit number, recognition of double-digit number recognition, subtraction of two-digit number with a carry over, and division of a three digit by one digit division. We use standardized scores of these variables as our outcome indicators.

One of the problems using ASER dataset poses is the paucity of data on cooking fuel usage at the household level. Accordingly, use district level share of rural households using solid fuel for cooking purpose from NFHS-4 and merge it with the ASER dataset at the district level conducted in 2016, the same year NFHS-4 was conducted. This district level solid fuel usage is used as the primary variable of interest along with a female dummy that takes the value of 1 if the child is female and 0 otherwise. This allows us to derive indicative evidence on the potential effect of solid fuel usage on learning outcomes among rural children.

The regression results are presented in figure 4. We run four regressions, out of which the first two indicate if children from districts with higher share of solid fuel usage score lesser in reading tests separated by gender and the next two present the effects for mathematics scores. We run two more regressions to understand if female children from districts with higher hare of solid fuel usage perform worse on average than the males. This is obtained by examining the marginal effects of the interaction term of the female dummy and the share of solid fuel usage at the district level. As one would expect, we observe females from these districts are more likely to fare worse than the males not only in reading but also in arithmetic outcomes. The lower dedicated time that children can give to schooling and homework because of solid fuel usage may manifest with lower mathematics abilities, where females on average find themselves in a disadvantageous position. Further, this finding is important as Choudhuri and Desai (2021) find detrimental mathematics score for males in their early childhood because of lesser time dedicated by mothers due to time substitution for firewood and water collection. Our finding actually suggests a reversal at an adolescent stage wherein the gender gap among females increase in regions with a higher prevalence of solid fuel being used as cooking fuel and possibly within households that use it as well. Nevertheless, further in depth study on the implications of household solid fuel on differentiated learning outcomes on this is required and our paper gives an indicative evidence of this, which can serve as a motivation for future research.

[Insert Figure 4 here]

### *5.4 Gendered effect for younger and older children*

Our results suggest that for households using solid fuel, the time substitution is stronger for adolescent females putting them at a more disadvantageous position relative to the males. If this is true, older adolescent females who are likely to allocate higher time for free collection of firewood and other resources may have to suffer disproportionately more in terms of educational outcomes. This comes from well-established literature especially in the developing country context on traditional household roles that older adolescent females are expected to perform (Edmonds, 2006; Kambhampati and Rajan, 2008). These among other things include domestic chores, care activities and collection of water or firewood. From the TUS data as well, we find the time allocated by older adolescent females (16 to 18 years) for collection of firewood among rural households using solid fuel for cooking purpose is significantly higher in comparison to the females in the lower age cohort and also boys in either of the two cohorts. An older adolescent female is likely to spent about an average 0.17 hours (95% CI: 0.15-0.19) as against 0.11 hours for the younger ones (95% CI: 0.1-0.13).[9] The corresponding figures for males are 0.065 (95% CI: 0.05-0.08) and 0.047 (95% CI: 0.036-0.058) respectively.

Accordingly, we examine if the relative position among females changes with age or remains stable over time. To assess this, as explained earlier, we split the sample into younger (12-15 years) and older cohort (16-18 years) children and then study the effect of solid fuel across gender. In addition to the three educational outcomes variables, we also consider primary school completion as an additional outcome for the younger cohort. Primary completion takes the value of 1 if the child has completed at least five years of schooling and 0 otherwise. Additionally, we analyze the effect of fuel use on secondary completion for children in the age group of 17-18 years since the outcome variable is relevant for children of this age group and not for the younger ones. Secondary completion takes the value of 1 if the child has completed at least 10 years of schooling and 0 otherwise. Figure 5 presents the second stage 2SLS marginal effects for these age groups.

[Insert Figure 5 here]

For the younger cohort, solid fuel is negatively related to school attendance with no apparent difference across gender; however, for children in the older cohort we do not find any discernible effect of household solid fuel usage on attendance in schools (Figure 5a). Further, females appear to be more adversely affected as the solid fuel bearing is found to be significantly negative for years of schooling while no such effects are observed for the males (Figure 5b). Next, we observe that younger females are likely to be marginally more affected when grade progression is progression is considered and this gender gap gets magnified in the older age group (Figure 5c). Finally, using solid fuel is found to reduce the likelihood of primary and secondary school completion for both males and females though the effect size is more prominent for females especially in the 17-18 years age-group (Figure 5d). Given that secondary education is more closely linked to future potential earnings, the striking effect of solid fuel use on secondary completion, grade progression and schooling for females compared to males contribute further to the gender gap in labor force participation as well as wages. Furthermore, lower educational attainment of girls on a whole is likely to fetch detrimental outcomes for females in terms of their relative bargaining, fertility decisions and autonomy among others. Because the gendered effect of solid fuel on educational outcomes of children gets magnified in the older age group, it can potentially have larger implications on labor market outcomes, marital decisions and empowerment.

[9] CI stands for Confidence Interval.

### *5.5 Health as an additional channel*

Extant literature documents adverse health impacts of solid fuel among children as well as for adults (Lin et al., 2008; Edwards and Langpap, 2012; Gupta, 2019; Liu et al., 2020). The negative effect of solid fuel use on years of schooling can hence be also driven by the poor health of adolescent children in addition to the time substitution channel that we emphasized in the earlier sections. Studies suggest that intricate health condition often lead to reduced school absenteeism and affects, school enrolment, literacy rate and lower academic achievement (Miguel and Kremer; 2004; Bleakley; 2007; Ding et al., 2009). Further, adverse health may have differential effects on males and females owing to gender discrimination. First, women and young adolescent females are predominantly involved in cooking and hence are more likely to be affected due to direct and longer exposure to smoke due to combustion of solid fuel (Chen and Modrek, 2018). Secondly, evidence suggests that adverse health shocks may affect the education of females more relative to males owing to gender bias in resource allocation (Maccini and Yang, 2009). In other words, given the relative lower returns to female education, parents may reallocate resources form education spending of female to meet health expenses. Finally, as females are more likely to be involved in care-giving for sick members of the households compared to males (Hirway and Jose, 2011), poor health of a household member owing to solid fuel use may have a larger detrimental effect on female's education.

The NFHS-4 survey does not provide information on health indicators of all the respondents but collects information on disease burden in addition to reporting the blood pressure and glucose level for males and females in the age group of 15-49years.[10] Hence, we are unable to explore the effect of solid fuel on health outcomes of adolescent children which in turn can potentially have a negative impact on education. However, the survey asks the question "whether any household member suffers from tuberculosis?"[11] We use this question to measure tuberculosis (TB) incidence at the household level and use it as a health indicator as studies document a positive correlation between solid fuel usage and tuberculosis (Lin et al., 2008; Kurmi et al., 2011; Popovic et al., 2019). The TB variable takes the value 1 if any household member suffers from TB and 0 otherwise. Notably, only 1.8% of the households reported someone suffering from TB. We estimate a household level regression of TB on solid fuel as the outcome variable is not at the child level. In this specification we exclude child characteristics but additionally control for sex of the household head and household head's age along with other household factors included in the baseline model. Columns 1of Table A7 of the appendix suggest that solid fuel usage is positively associated with the likelihood of TBat 1% level of significance, however, in IV result reported in column 2, the coefficient of solid fuel is positive but no longer significant (p-value: 0.2). This could be because of the skewed distribution of TB variable. However, there appears to a positive albeit weak correlation between TB and solid fuel use as suggested in the literature. . Next, to establish TB as a possible channel, following Alesina and Zhuravskaya (2011) and Zhang et al. (2021), we re-estimate the effect of solid fuel on educational outcomes for males and females separately after including TB as an additional control. The IV results presented in columns 3-8 indicate that the coefficient of TB is negatively related to educational outcomes for both males and females and the coefficient of solid fuel in these specifications is marginally lower than those presented in Figure 2. The evidence does not allow us to rule out health as a channel through which solid fuel leads to inferior educational outcomes. We also do not observe any noteworthy difference across males and females. Consequently, whether poor health due to solid fuel usage contributes to the gendered effect on education remains inconclusive and is an area for future research.

[10] The survey collects biomarkers and other health outcomes for men and women in the age-group of 15-54 years and 15-49 years, respectively.
[11] Survey also captures whether the individual suffers from tuberculosis, since less than 0.1% children have tuberculosis we do not consider this.

## 6. Discussion in the light of the COVID-19 pandemic

The findings from the paper which establish the associated ill effects of solid fuel usage on human capital investment, which appears to be disproportionately more adverse for females and calls for actions to ensure faster transition to cleaner fuel options. The understanding and the need for higher adoption of cleaner fuel and its effect of education becomes even more pertinent in the present scenario as the outbreak of the COVID-19 across the world may not only reduce the pace of clean fuel adoption, but also threatens to reverse the initial gains in energy transition. Recent studies in the context of Africa find that the pandemic induced income shocks led to substitution of LPG with cheaper polluting fuels like coal and kerosene by households (Shupler et al., 2021a). Given that the use of solid fuel is mostly concentrated in the Global South, any further movement down the energy ladder can potentially widen the energy gap between the low and middle income and high income countries apart from increasing inequality between rich and poorer within these developing economies. Further, studies find that job loss in paid work was higher for women compared to men (Desai et al., 2021; Dang and Nguyen, 2021). This gendered pattern of unemployment may reduce women's intra-household bargaining power and households may end up reducing expenditure on cooking fuel (an activity typically performed by women) by shifting to free collection of firewood.

The COVID-19 induced lockdown and school closures have already aggravated the educational gap between the rich and the poor[12] and it may further widen the educational gap between girls and boys (Alvi and Gupta, 2020). Parental job loss along with social norms and gender digital divide can also potentially lead to higher school dropout among children especially for adolescent girls as limited financial resources are more likely to be redirected towards human capital investments for boys (World Bank, 2020). Further, due to the pandemic, the increased domestic unpaid work and care work has disproportionately befallen on women and girls (Kabeer et al., 2021; Seck et al., 2021; Desai et al., 2021; World Bank, 2020). The higher demand for care tasks along with the shift to solid fuel may further endanger continuation of girl's education. The gendered effect of the pandemic on females may reverse the fragile gains in reducing gender inequality in education achieved in past few decades.

Further, at a macroeconomic level, the pandemic led to budget tightening, especially in the worst-hit countries including India (Hosseini, 2020) and the risk of budgetary reallocation from clean fuel subsidies to more pressing demands looms large. This along with a significant fall in global fossil fuel prices especially during 2020 can possibly become a major hurdle for clean energy adoption as the relative price of using clean energy sources have gone up significantly (Hoang, 2021). While Smith et al. (2021) argue that the fall in the price of fossil fuel and the COVID-19 pandemic may not affect the climate change mitigation efforts by the countries, both macro and microeconomic evidence largely indicates that the pandemic poses a real threat of delaying in the energy transition to cleaner fuel in the Global South (Ravindra et al. 2021; Shupler, 2021a). However, it is worth noting that a recent study by Shupler et al. (2021b) find that the pay as you go LPG users in Kenya were less likely to reduce cooking time during lockdown than conventional LPG cylinder users highlighting the need to deploy technology based solutions to ensure that the fragile gains towards clean energy adoption do not get reversed and in fact the household energy transition trajectory gets accelerated in the post-covid recovery period.

[12] https://sdgs.un.org/goals/goal4 (accessed on May 31, 2021).

## 7. Conclusion

Reliance on solid fuel for cooking and its adverse impact on health because of the pollutants emitted by the fuel is well documented in the literature. This paper complements this literature on welfare reducing impact of solid fuel use by exploring its effects on human capital investments among rural adolescent children from India. Our findings indicate a robust and significant deterioration in educational outcomes among children from households using solid fuel. We find these effects not only on school attendance but also on years of education and age appropriate grade progression. Importantly, we are able to account for the potential bias arising because of self-selection and this allows us to draw causal inference from our regression estimates. Additionally, we find a significant female disadvantage with the females more likely to suffer in terms of the educational outcomes. While we do not rule out the possibility of these effects emanating indirectly from the adverse impact on health, we consider the direct time substitution due to the free collection of firewood and other natural resources as a significant channel. We show that time spent in schools decreases significantly as time allocation for environmental chores like the collection of fuel resources increase in households using solid fuel for cooking. We further find that the extent of the substitution to be disproportionately higher for females. In fact, among the adolescent children, the gender difference is found to be marginal for younger age cohort but this gap diverges for the older group indicating that the elder adolescent females bear most of the brunt. In this connection, we also find that the likelihood of secondary school completion to be lower, which is observed to be causally related to cooking fuel choice.

Despite important and policy relevant findings, our study has limitations and hence opens up possibilities for future research. Firstly, in the absence of recent longitudinal data at the household level, we are unable to track the changes in educational outcomes with transition to cleaner fuel. This would have allowed us to obtain more precise causal estimates. Secondly, as mentioned, the ASER dataset that we use does not gather information on the cooking fuel used by the surveyed households. Therefore, we are able to present only suggestive evidence linking district level prevalence of solid fuel usage with learning outcomes among children. Our study gives the necessary motivation to study the relationship between the two using household level cooking fuel choice and learning abilities among the children as future research. Finally, while we are able to give satisfactory evidence of time allocation for natural resources directly hindering educational outcomes in households using solid fuel, we are not able to reject the indirect role of adverse health outcomes affecting the gendered pattern of the effect on schooling and educational attainment because of paucity of relevant data. This as well can be considered as one for further research.

Despite these limitations, our paper has important policy implications. The findings underscore the importance of the need for policies that encourage household adoption to LPG or other cleaner options. Apart from the detrimental effects on health and environment, our study finds complementary evidence of adverse effects on human capital investments. This has huge relevance as literature has established the very significant role played by education on labor market opportunities and earnings (Kingdon, 1998; Jensen, 2010). This is particularly important in a country like India with evidence to suggest that a considerable proportion of population not having enough education and skills and thus reducing their employability (Blom and Saeki, 2011; Unni, 2016). This is even true at lower levels of education as a study conducted in 2018 found about half of the rural children in India studying in fifth grade are unable to read texts meant for students studying in the second grade and also solve two-digit subtraction problems (Pratham, 2019). From the perspective of rural India, literature has documented substantial income gains in the agrarian sector linked with higher education (Duraisamy, 1992). Accordingly, incentivizing households to move up the energy ladder and adoption of cleaner fuel in rural India assumes significance due to positive spillovers on education in addition to health concerns.

The paper also presents evidence of an educational disadvantage, which is disproportionately higher for females especially at the later period of adolescence. This is pertinent in the context of India which already suffers from low female labor participation (Chowdhury, 2011; Neff et al., 2012). Apart from this, it is also found to bestow a number of non-market benefits especially for females that include among others, marital decisions, fertility choices and empowerment emanating through higher relative bargaining power because of education and alteration of gender norms. Consequently, to ensure better opportunities for females and reduce gender inequality, focus on the adoption of cleaner fuels for cooking purposes becomes imminent.

A number of interventions have been initiated in India to increase the use of LPG that includes the *Rajiv Gandhi Gramin LPG Vitaran Yojana (RGGLVY)* in 2009 and the *Pratyaksh Hastantrit Labh (PAHAL)* in 2015. More recently, the *Pradhan Mantri Ujjwala Yojana* (PMUY) has been implemented since 2016. Under the provisions of the PMUY, LPG connections are provided to poor households in the name of a woman household member who is above 18 years of age. While there has been a considerable expansion of coverage and distribution network over the recent years with evidence of a socially inclusive transition to cleaner fuel, its usage over time does not yield encouraging feedback (Swain and Mishra, 2020). Apart from issues with targeting, studies have indicated higher prices discourage LPG refills among the beneficiaries of the program and hence they often transition back to solid fuels (Kar et al. 2019; Gould and Urpelainen, 2020; Swain and Mishra, 2020). Additionally, subsidization of alternative fuel like kerosene or lower transaction cost of accessing biomass and firewood, increases the possibility of crowding out of the more expensive fuels like LPG. This underscores the need for policy interventions that call for higher and regular subsidization of cleaner fuel in the short run to reduce fuel stacking by rural households. Further, cylinders fitted with smart meters that allows households to partially refill them using mobile also popularly known as pay as go LPG (PAYGL) cylinders may fasten the adoption and use of clean fuel in rural households reducing their financial burden. The PAYGL is already implemented and has been received well by a few East African economies.[13] Additionally, awareness programs regarding the negative impacts of dirty fuel and behavioral intervention to address the cultural believes surrounding cooking practices that favor solid fuel (Martinez et al., 2020; Williams et al., 2020) has the potential to improve clean fuel adoption and consistent usage in the longer run.

[13] https://www.paygoenergy.co/about (accessed on May 31, 2021).

**Figures**

Figure 1: Schooling for solid and non-solid fuel users

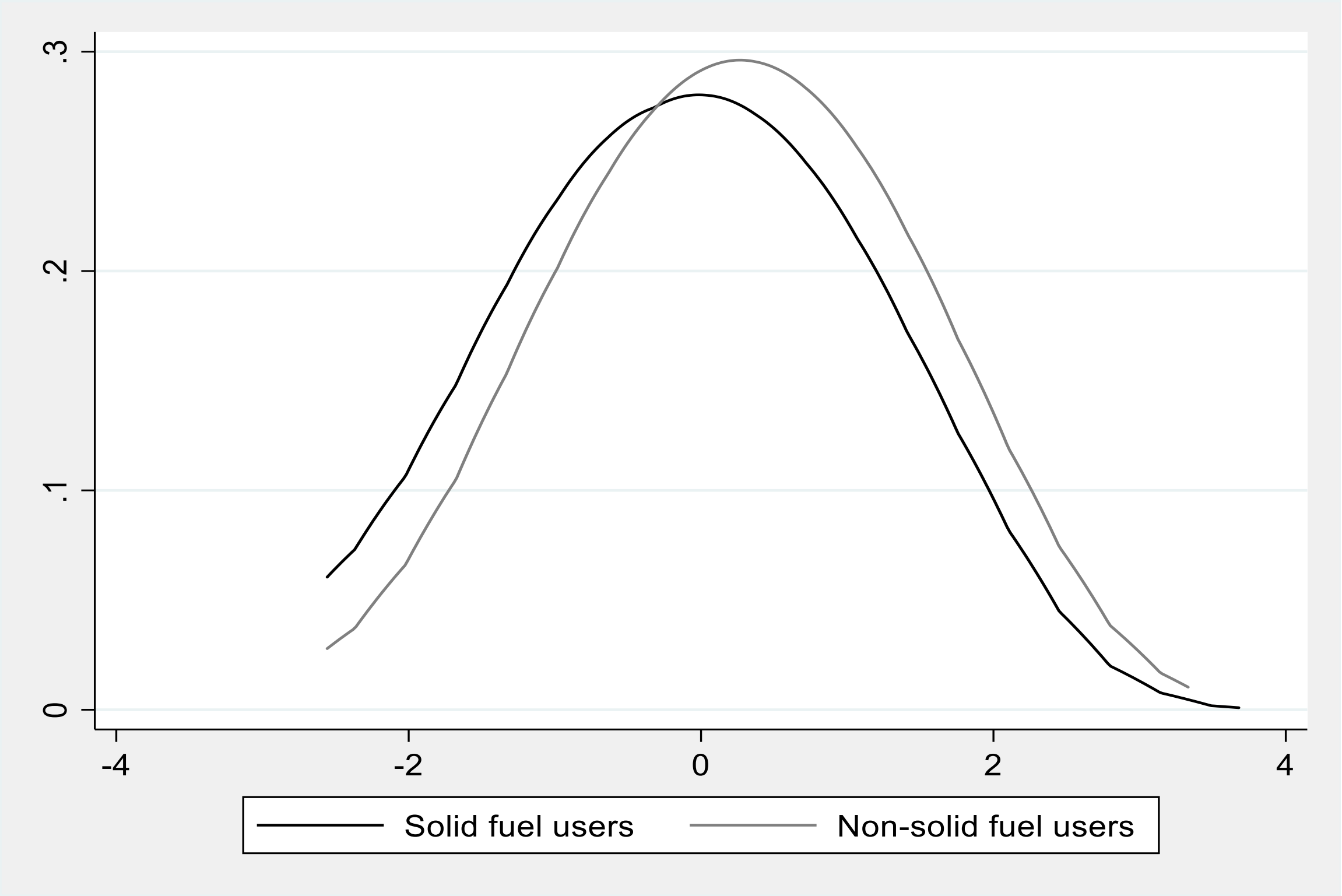

Note: Figure presents the kernel density of years of education (standardized) of rural children from 12 to 18 years of age from solid fuel and non-solid fuel user households.

Figure 2: Effect of solid fuel on educational outcomes – Males and females

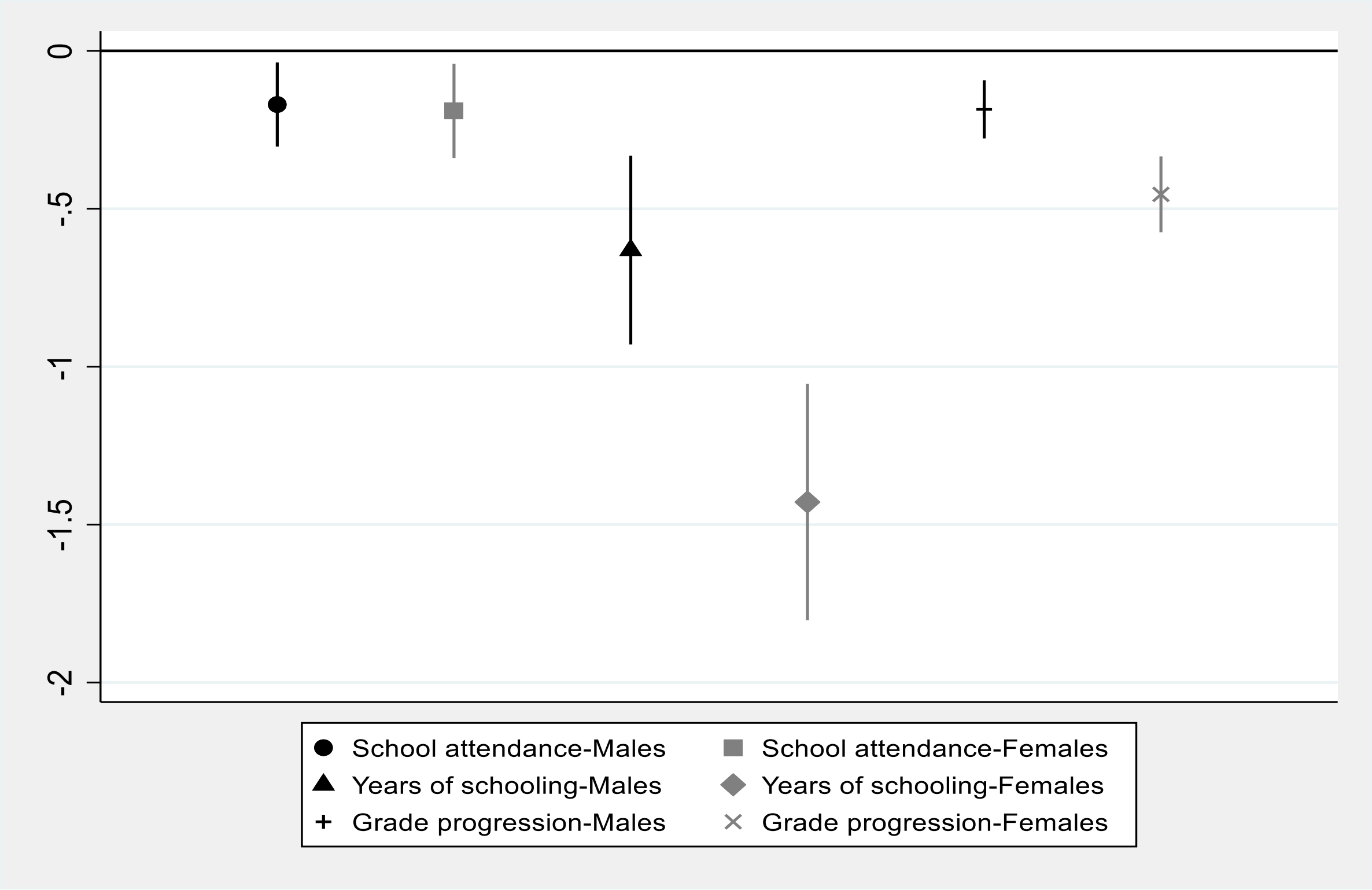

The average marginal effects are plotted along with the 90% confidence intervals. All regressions control for the covariates including current age, number of primary completed adults, household size, BPL card, toilet, water in yard, mobile, upper caste, religion, month of survey and state fixed effects. The sample is rural children from 12 to 18 years of age. Regression output is based on robust standard errors.

Figure 3: Effect of time taken for firewood and other natural resource collection on school and homework time

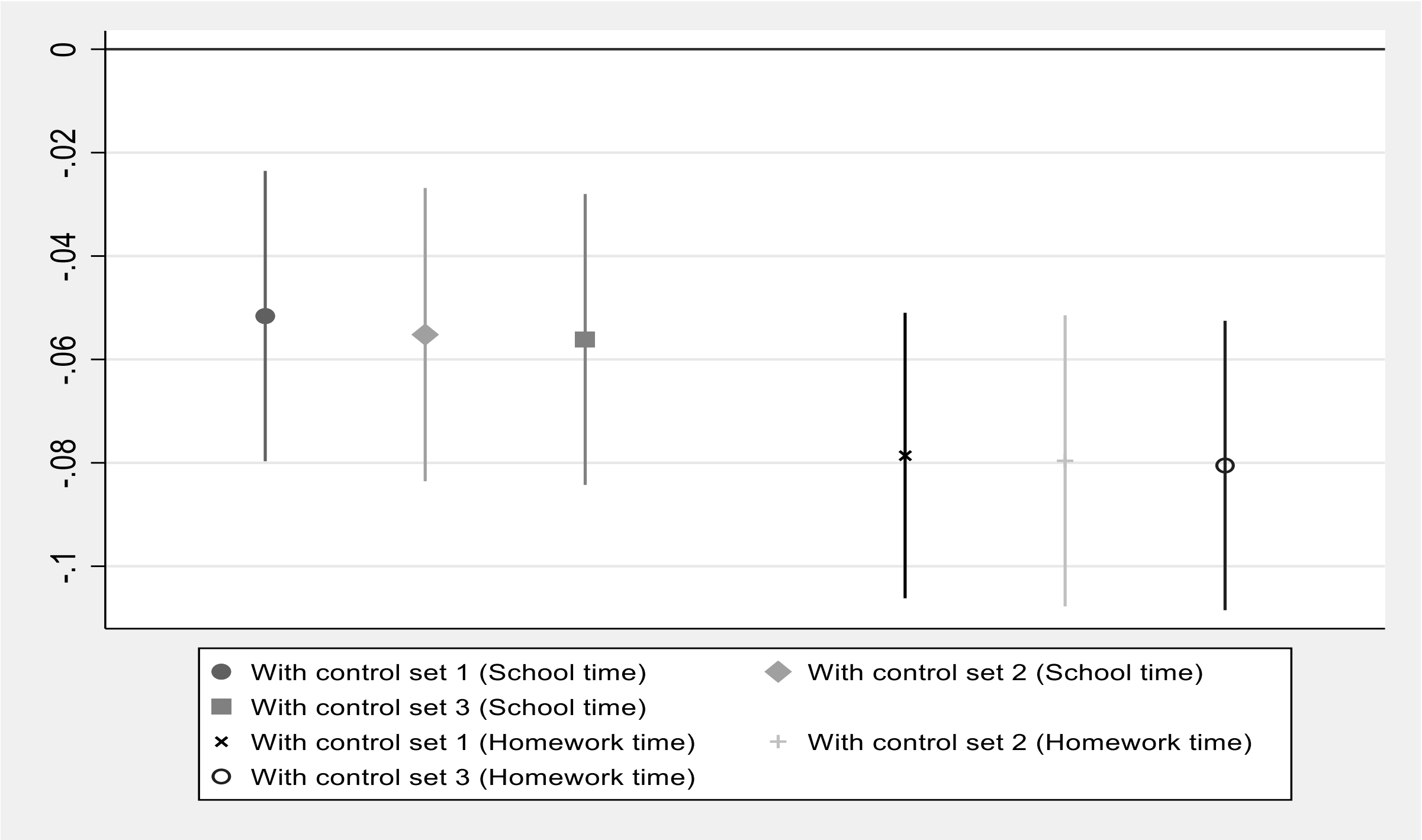

Note: The average marginal effects are plotted along with the 90% confidence intervals. Control set 1 includes current age, number of primary completed adults, household size, usual monthly household consumption expenditure (in INR), upper caste, religion, month of survey and state fixed effects. Control set 2 includes all the above as well as the time allocated by others in the household for firewood and other natural resource. Control set 3 includes all those included in control set 2 along with the time allocated for care giving. The sample is rural children from 12 to 18 years of age. Regression output is based on robust standard errors.

Figure 4: Association of district level solid fuel use and performance using ASER data

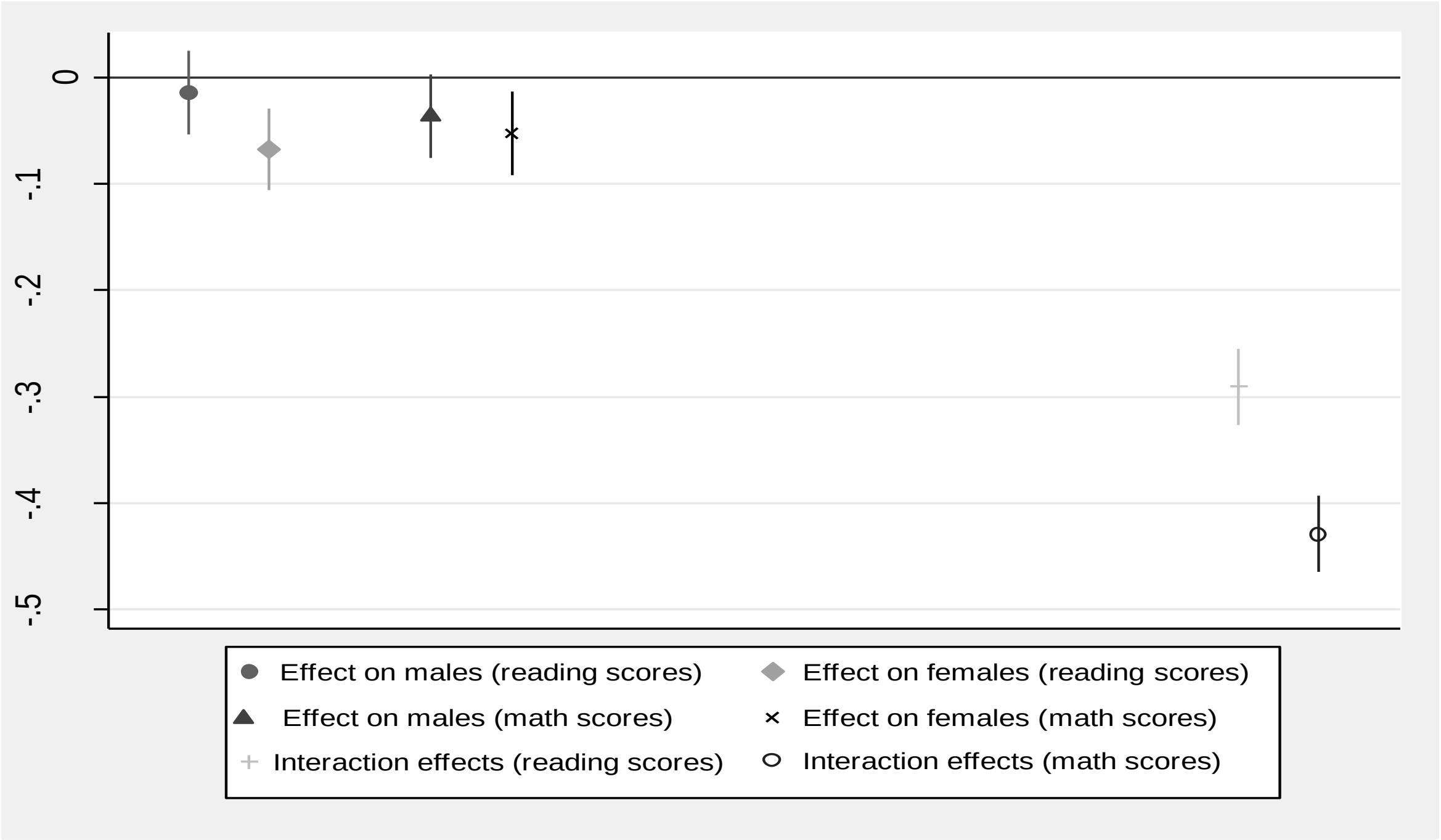

Note: The average marginal effects are plotted along with the 90% confidence intervals. All regressions control for the covariates including current age, household possession of television and toilet, household size, whether the household is cemented or not, whether anyone in the household knows computer usage, whether anyone in the household has a mobile phone, whether the mother attended school or not and state fixed effects. The sample is rural children from 12 to 16 years of age. Regression output is based on robust standard errors.

Figure 5: Age-cohort wise effect on educational outcomes for males and females

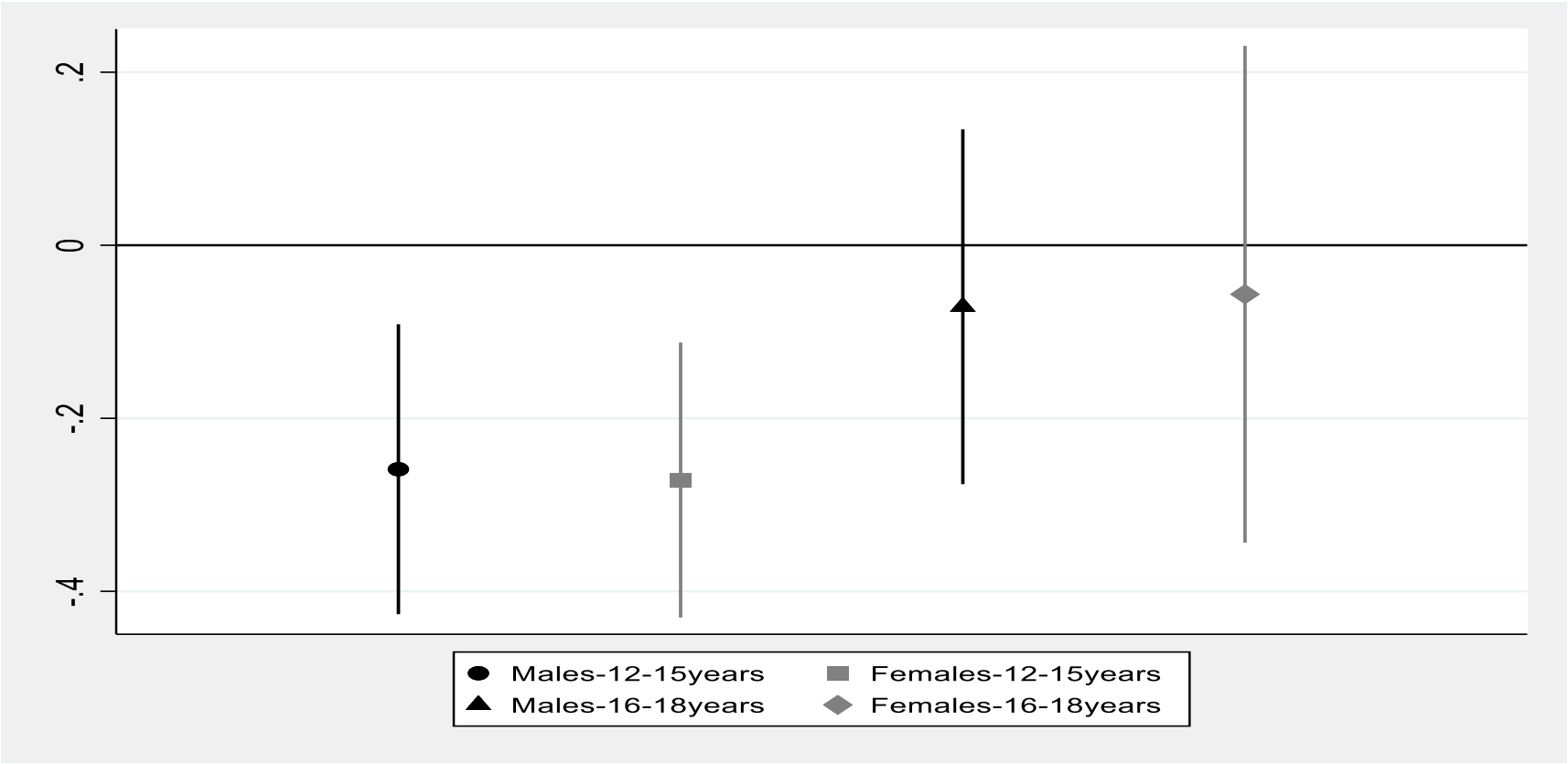

5a: School attendance

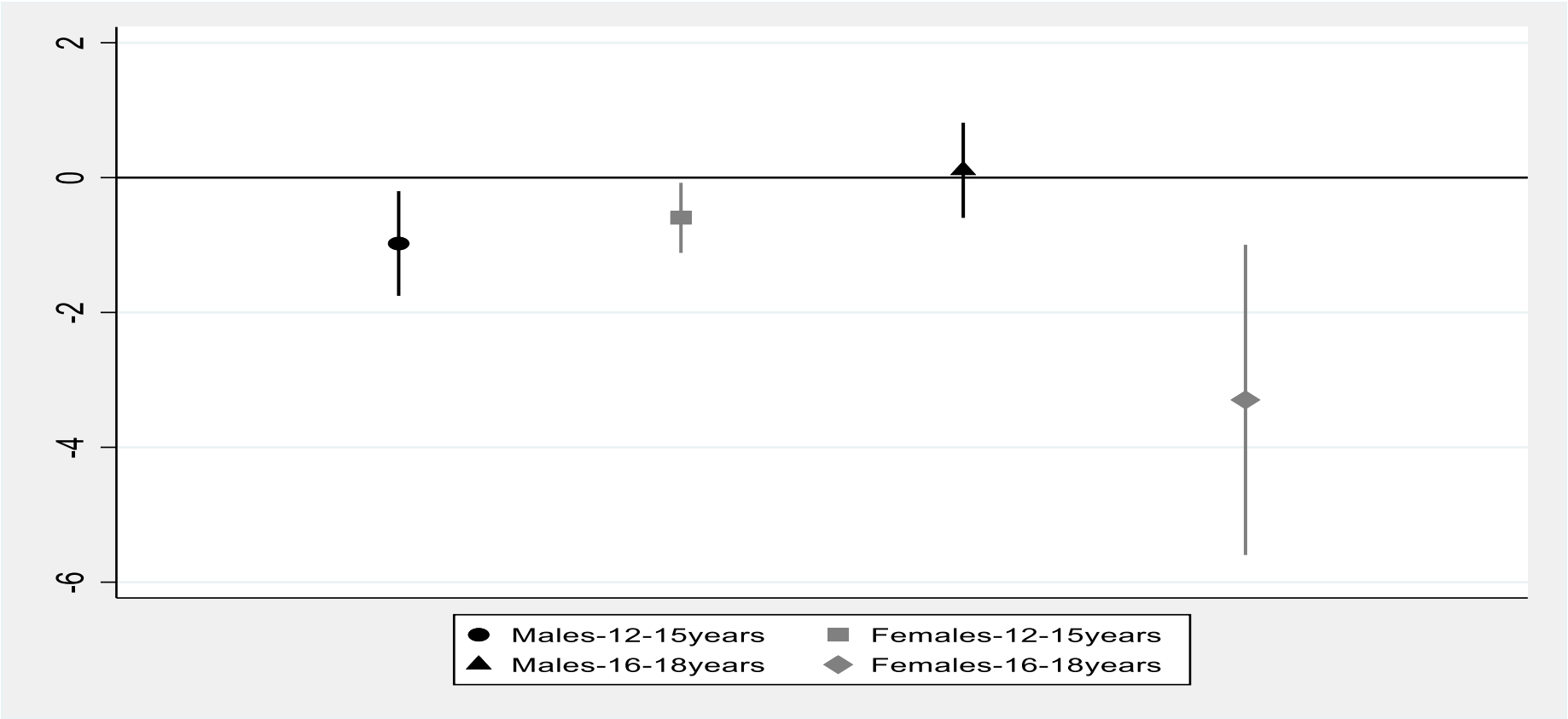

5b: Years of schooling

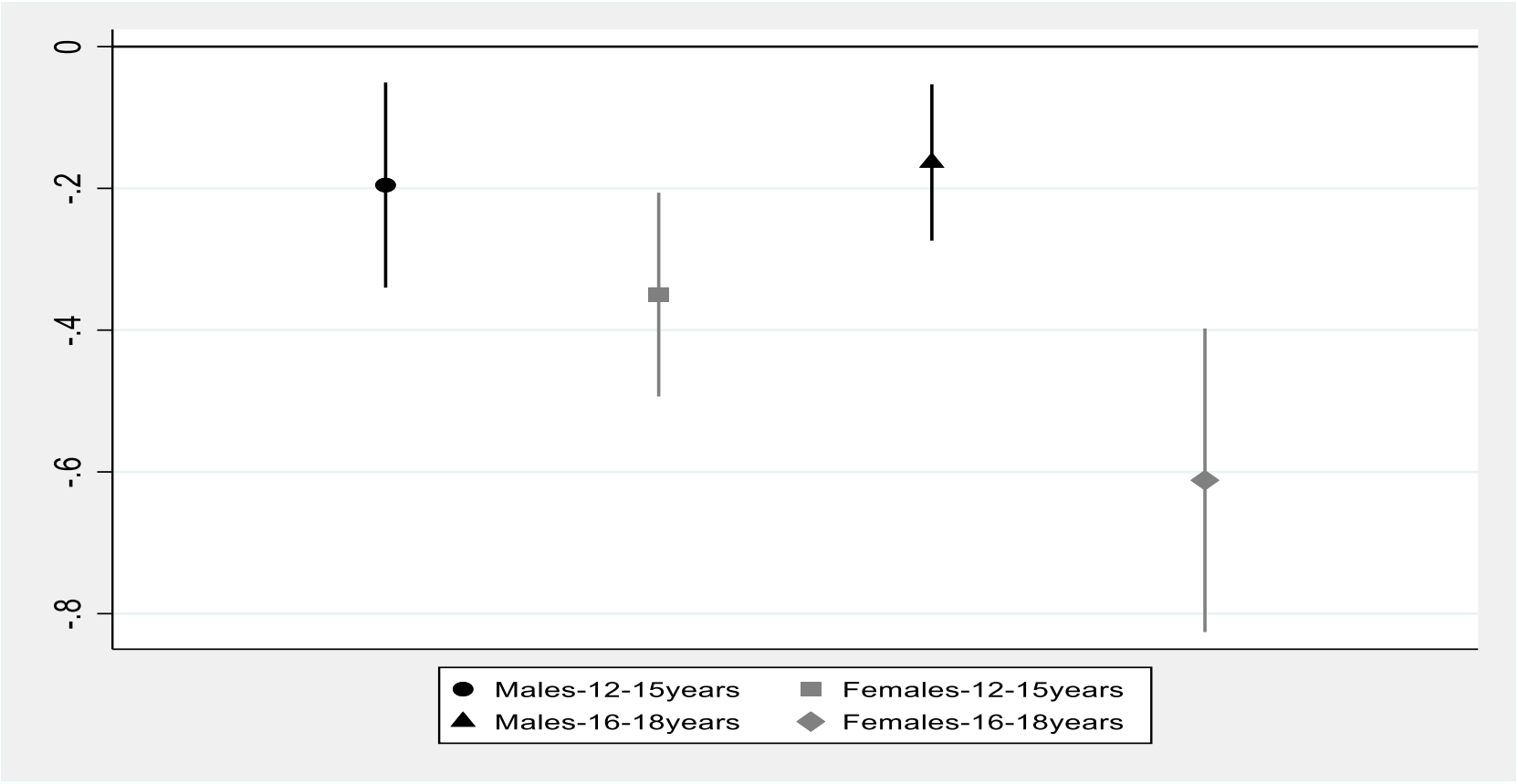

5c: Grade progression

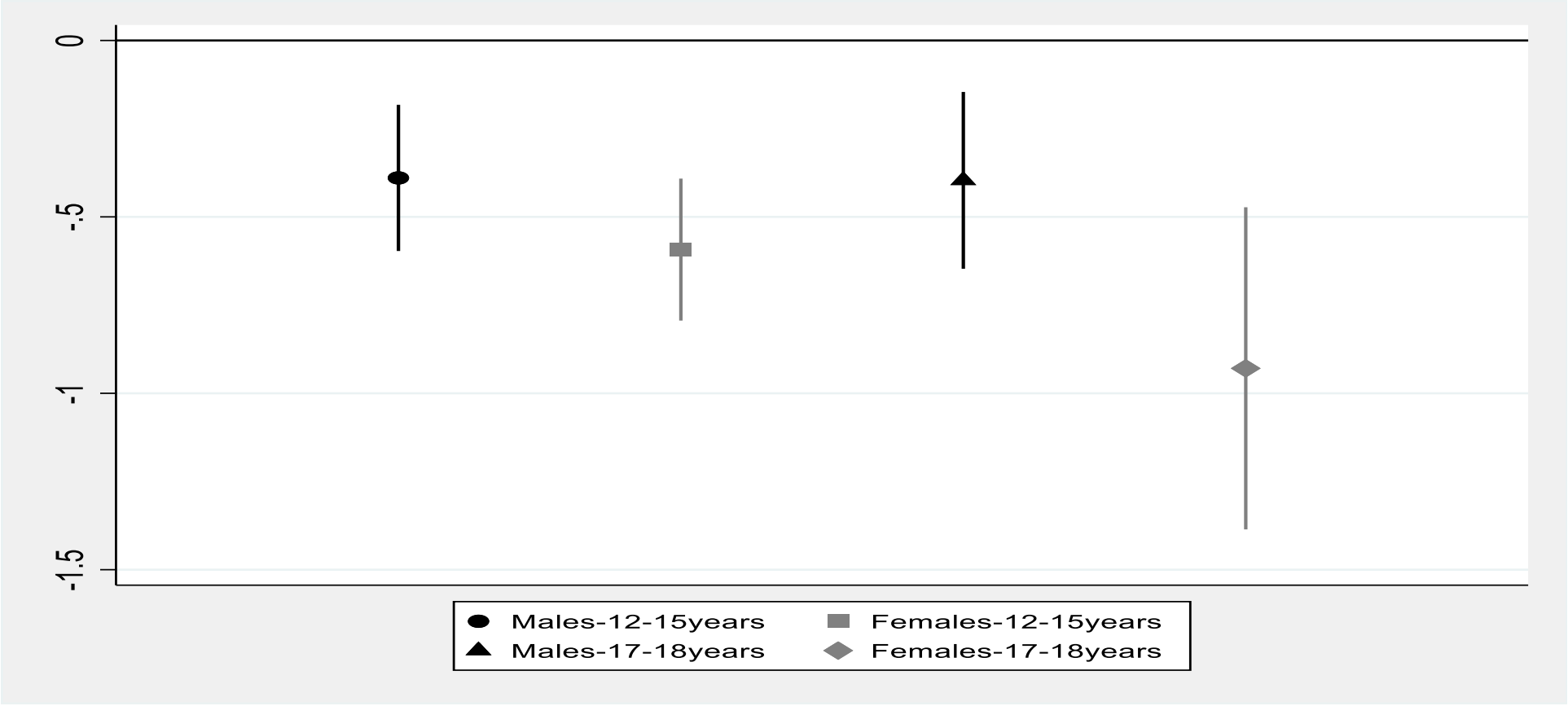

5d: Primary and secondary completion

Note: The average marginal effects are plotted along with the 90% confidence intervals. All regressions control for the covariates including current age, number of primary completed adults, household size, BPL card, toilet, water in yard, mobile, upper caste, religion, month of survey and state fixed effects. The sample is rural children from 12 to 18 years of age. Regression output is based on robust standard errors.

## Tables

Table 1: Variable definitions

| Variable | Description |
|---|---|
| *NFHS-4 (2015-16)* | |
| ***Outcome variables*** | |
| **School attendance** | Dummy variable=1 for children who attended school during the survey year and 0 for others |
| **Years of schooling** | Standardized years of schooling |
| **Grade progression** | Ratio of actual years of schooling and expected years of schooling given age. Expected years of schooling=Current age- 6years |
| **Primary** | Dummy variable=1 for individuals who have completed at least 5 years of schooling and 0 for |
| **Secondary** | Dummy variable=1 for individuals who have at least 10 years of schooling and 0 for others |
| ***Interest variable*** | |
| **Solid fuel** | Dummy=1 for households using firewood, animal dung, agricultural crop, straw/shrub/grass, coal/lignite or charcoal as cooking fuel and 0 if cooking fuel used is LPG, electricity, biogas or kerosene |
| **Dirty fuel** | Dummy=1 for households using firewood, animal residue, crops or kerosene as cooking fuel and 0 if cooking fuel is LPG, electricity or biogas |
| **District level solid fuel usage** | Proportion of households using firewood, animal dung, agricultural crop, straw/shrub/grass, coal/lignite or charcoal as cooking fuel in the district |
| ***Instrument*** | |
| **Average forest cover** | Average share of forest in the district 6 years before the survey year. |
| ***Other controls*** | |
| **Age** | Age of the child in years |
| **Female** | Dummy=1 for girls and 0 for boys |
| **Primary completed adults** | Number of household members above 18 who have completed primary education |
| **Household size** | Number of household members |
| **BPL card** | Dummy=1 if the household has a BPL card and 0 otherwise |
| **Toilet** | Dummy=1 if the household has toilet facility in within premises and 0 otherwise |
| **Water in home/ yard** | Dummy=1 if the household has access to piped drinking water in the house or yard and 0 otherwise |
| **Mobile** | Dummy=1 if any member in the household owns a mobile and 0 otherwise |
| **Religion** | Categorical variable identifying the religious affiliation of the individual as Hindu, Muslim, Others |
| **Upper caste** | Dummy=1 if the household does not belong to Scheduled castes, Scheduled Tribe or Other Backward |
| *TUS (2019)* | |
| *Outcome variables* | |
| **Total school time (standardized)** | Time allocated in hours (standardized) for School/university attendance, Extra-curricular activities, breaks at place of formal education, self-study for distance education course work, other activities related to formal education |
| **Total homework time (standardized)** | Time allocated in hours (standardized) for homework, being tutored, course review, research and activities |
| *Interest variable* | |
| **Total fuel collection time** | Time allocated in hours for Gathering firewood and other natural products used as fuel for own final use |
| **Solid fuel usage** | Dummy=1 for households using firewood and chips, dung cake, coke or coal, or charcoal as primary source of energy for cooking during last 30 days preceding the date of survey and 0 if cooking fuel used is LPG, other natural gas, electricity, gobar gas, other biogas, |
| *Instrument* | |
| **Average forest cover** | Average share of forest in the district 6 years before the survey year (2013). |
| *Other controls* | |
| **Age** | Age of the child in years |
| **Female** | Dummy=1 for girls and 0 for boys |
| **Primary completed adults** | Number of household members above 18 who have completed primary education |
| **Household size** | Number of household members |
| **Household consumption expenditure** | Usual monthly consumer expenditure (in Indian rupees- INR) for the household |
| **Religion** | Categorical variable identifying the religious affiliation of the individual as Hindu, Muslim, Others |
| **Upper caste** | Dummy=1 if the household does not belong to Scheduled castes, Scheduled Tribe or Other Backward caste and 0 otherwise |
| *ASER (2016)* | |
| *Outcome variables* | |
| **Reading scores (standardized)** | Reading scores are as follows: 1: Unable to read; 2: Can identify letters; 3: can read a word; 4: can read a paragraph; 5: can read a story |
| **Mathematics scores (standardized)** | Mathematics scores are as follows: 1: Unable to do any math; 2: Can identify numbers 1-9; 3: can identify numbers 10-99; 4: can do a subtraction; 5: can do a division. |
| *Controls* | |
| **Age** | Age of the child in years |
| **Female** | Dummy=1 for girls and 0 for boys |
| **Cemented house** | Dummy=1 for cemented households (with walls and roofs made of brick and cement; 0 otherwise |
| **Household toilet** | Dummy=1 if there is a toilet in the household; 0 otherwise |
| **Television** | Dummy=1 if there is a television in the household; 0 otherwise |
| **Computer usage** | Dummy=1 if anyone in the household knows how to use computers and 0 otherwise |
| **Mobile** | Dummy=1 if anyone in the household has a mobile phone; 0 otherwise |
| **Household size** | Number of household members |
| **Mother went to school** | Dummy=1 if the mother ever attended school; 0 otherwise |

Table 2: Summary statistics

| Variables | All households | Non-solid fuel users (a) | Observations | Solid fuel users (b) | Observations | Mean Difference (a-b) |
|---|---|---|---|---|---|---|
| School attendance (share) | 0.777 | 0.892 | 51563 | 0.752 | 242684 | 0.140*** |
| Years of schooling (in years) | 7.382 | 0.939 | 51511 | 0.815 | 242386 | 0.123*** |
| Grade progression (ratio) | 0.837 | 8.387 | 51511 | 7.165 | 242386 | 1.222*** |
| Age (in years) | 14.921 | 14.994 | 51569 | 14.905 | 242731 | 0.089*** |
| Female (share) | 0.487 | 0.478 | 51569 | 0.489 | 242731 | -0.011*** |
| Primary completed adults | 1.527 | 2.111 | 51569 | 1.403 | 242731 | 0.708*** |
| Household size | 6.165 | 5.655 | 51569 | 6.273 | 242731 | -0.619*** |
| BPL (share) | 0.472 | 0.323 | 51431 | 0.503 | 242444 | -0.180*** |
| Upper caste (share) | 0.17 | 0.271 | 48064 | 0.149 | 232897 | 0.122*** |
| Toilet (share) | 0.467 | 0.804 | 51569 | 0.395 | 242731 | 0.409*** |
| Water in yard (share) | 0.186 | 0.367 | 51569 | 0.147 | 242731 | 0.219*** |
| Mobile (share) | 0.914 | 0.982 | 51569 | 0.899 | 242731 | 0.083*** |
| Hindu | 0.751 | 0.681 | 51519 | 0.766 | 242585 | -0.085*** |
| Muslim | 0.131 | 0.154 | 51519 | 0.126 | 242585 | 0.028*** |
| Others | 0.118 | 0.166 | 51519 | 0.108 | 242585 | 0.057*** |

Note: Standard t-test is used to compare the difference in group means. *indicates significance at 1% level. The sample is rural children from 12 to 18 years of age.

Table 3: Effect of solid fuel on educational outcomes

| | (1)<br>School attendance-Probit | (2)<br>Years of schooling-OLS | (3)<br>Grade progression-OLS | (4)<br>School attendance-IV | (5)<br>Years of schooling-IV | (6)<br>Grade progression-IV |
|---|---|---|---|---|---|---|
| Solid fuel | -0.060*** | -0.066*** | -0.019*** | -0.182*** | -1.011*** | -0.311*** |
| | (0.002) | (0.004) | (0.001) | (0.061) | (0.142) | (0.045) |
| Other controls | Y | Y | Y | Y | Y | Y |
| State FE | Y | Y | Y | Y | Y | Y |
| Month of survey FE | Y | Y | Y | Y | Y | Y |
| First stage F-stat | | | | 253.250 | 253.250 | 253.250 |
| Observations | 280,339 | 280,011 | 280,011 | 280,339 | 280,011 | 280,011 |
| R-squared | | 0.346 | 0.206 | 0.187 | 0.247 | 0.092 |

Note: The regression coefficients are the average marginal effects. All regressions control for the covariates including current age, number of primary completed adults, household size, BPL card, toilet, water in yard, mobile, upper caste, religion, month of survey and state fixed effects. The sample is rural children from 12 to 18 years of age. Robust standard errors in parentheses. *** $p<0.01$, ** $p<0.05$, * $p<0.1$.

Table 4: Plausibly exogenous instrument regressions

| | (1)<br>School attendance | (2)<br>Years of schooling | (3)<br>Grade progression |
|---|---|---|---|
| $\hat{\gamma}$ | -0.042*** | -0.235*** | -0.072*** |
| | (0.014) | (0.030) | (0.010) |
| Controls | Y | Y | Y |
| State FE | Y | Y | Y |
| Month of survey FE | Y | Y | Y |
| Observations | 280,375 | 280,047 | 280,047 |
| R-squared | 0.200 | 0.345 | 0.256 |
| $\hat{\beta}$ (UB) | 0.115 | 0.246 | 0.076 |
| $\hat{\beta}$ (LB) | -0.301 | -1.289 | -0.389 |
| $\gamma_{max}$ | -0.015 | -0.177 | -0.054 |

Note: All regressions control for the covariates including current age, number of primary completed adults, household size, BPL card, toilet, water in yard, mobile, upper caste, religion, month of survey and state fixed effects. The sample is rural children from 12 to 18 years of age. Robust standard errors in parentheses. *** p<0.01, ** p<0.05, * p<0.1.

Table 5: Propensity score matching

| Dependent variable | (1) Treated | (2) Control | (3) Difference | (4) T-statistic | (5) Rosenbaum bounds |
|---|---|---|---|---|---|
| School attendance | 0.753 | 0.809 | -0.056 | -9.23*** | 1.38-1.39 |
| Years of schooling | -0.071 | 0.082 | -0.153 | -9.36*** | 1.28-1.29 |
| Grade progression | 0.816 | 0.859 | -0.043 | -10.44*** | 1.29-1.30 |

Note: Matching is performed after controlling for covariates including current age, number of primary completed adults, household size, BPL card, toilet, water in yard, mobile, upper caste, religion, districts, month of survey and state fixed effects. The sample is rural children from 12 to 18 years of age.

## Appendix

Table A1: Effect of solid fuel on education- First stage regression and IV-probit

| | (1)<br>First stage- LPM | (2)<br>School attendance-IV Probit |
|---|---|---|
| Solid fuel | | -0.846*** |
| | | (0.263) |
| Average forest cover | 0.233*** | |
| | (0.015) | |
| Other controls | Y | Y |
| State FE | Y | Y |
| Month of survey FE | Y | Y |
| Wald test of exogeneity | | 4.67** |
| Observations | 280,339 | 280,339 |

Note: All regressions control for the covariates including current age, number of primary completed adults, household size, BPL card, toilet, water in yard, mobile, upper caste, religion, month of survey and state fixed effects. The sample is rural children from 12 to 18 years of age. Robust standard errors in parentheses. *** p<0.01, ** p<0.05, * p<0.1.

Table A2: Effect of solid fuel on educational outcome- Full table

| | (1) School attendance-Probit | (2) Years of schooling-OLS | (3) Grade progression-OLS | (4) School attendance-IV | (5) Years of schooling-IV | (6) Grade progression-IV |
|---|---|---|---|---|---|---|
| Solid fuel | -0.060*** | -0.066*** | -0.019*** | -0.182*** | -1.011*** | -0.311*** |
| | (0.002) | (0.004) | (0.001) | (0.061) | (0.142) | (0.045) |
| Age | -0.068*** | 0.200*** | -0.027*** | -0.070*** | 0.202*** | -0.026*** |
| | (0.000) | (0.001) | (0.000) | (0.000) | (0.001) | (0.000) |
| Female | -0.045*** | -0.013*** | -0.000 | -0.047*** | -0.016*** | -0.001 |
| | (0.001) | (0.003) | (0.001) | (0.001) | (0.003) | (0.001) |
| Primary completed adults | 0.061*** | 0.179*** | 0.056*** | 0.054*** | 0.139*** | 0.044*** |
| | (0.001) | (0.001) | (0.000) | (0.003) | (0.006) | (0.002) |
| Water in home/ yard | 0.017*** | 0.028*** | 0.009*** | 0.002 | -0.052*** | -0.016*** |
| | (0.002) | (0.004) | (0.001) | (0.006) | (0.013) | (0.004) |
| Toilet | 0.059*** | 0.164*** | 0.051*** | 0.032*** | -0.011 | -0.003 |
| | (0.002) | (0.004) | (0.001) | (0.011) | (0.026) | (0.008) |
| Own mobile | 0.076*** | 0.266*** | 0.088*** | 0.077*** | 0.219*** | 0.074*** |
| | (0.003) | (0.007) | (0.002) | (0.004) | (0.010) | (0.003) |
| Household size | -0.018*** | -0.070*** | -0.022*** | -0.017*** | -0.050*** | -0.016*** |
| | (0.000) | (0.001) | (0.000) | (0.001) | (0.003) | (0.001) |
| BPL card | -0.019*** | -0.045*** | -0.013*** | -0.011*** | 0.005 | 0.002 |
| | (0.002) | (0.003) | (0.001) | (0.004) | (0.008) | (0.003) |
| Upper caste | 0.028*** | 0.062*** | 0.019*** | 0.012*** | 0.016* | 0.005* |
| | (0.002) | (0.004) | (0.001) | (0.004) | (0.008) | (0.003) |
| Muslims | -0.148*** | -0.380*** | -0.121*** | -0.145*** | -0.394*** | -0.126*** |
| | (0.003) | (0.007) | (0.002) | (0.003) | (0.007) | (0.002) |
| Other religion | -0.007* | -0.112*** | -0.039*** | -0.009*** | -0.116*** | -0.040*** |
| | (0.004) | (0.007) | (0.002) | (0.003) | (0.008) | (0.002) |
| Constant | | -2.859*** | 1.269*** | 1.945*** | -2.233*** | 1.462*** |
| | | (0.026) | (0.008) | (0.042) | (0.098) | (0.031) |
| State FE | Y | Y | Y | Y | Y | Y |
| Month of survey FE | Y | Y | Y | Y | Y | Y |
| Observations | 280,339 | 280,011 | 280,011 | 280,339 | 280,011 | 280,011 |
| R-squared | | 0.346 | 0.206 | 0.187 | 0.247 | 0.092 |

Note: Robust standard errors in parentheses. *** p<0.01, ** p<0.05, * p<0.1.

Table A3: Solid fuel use and school timing using TUS

| | (1)<br>OLS | (2)<br>IV |
|---|---|---|
| Solid fuel | -0.088*** | -1.618** |
| | (0.010) | (0.670) |
| Other controls | Y | Y |
| Month of survey FE | Y | Y |
| State FE | Y | Y |
| F statistic | | 15.092 |
| Observations | 36,723 | 36,570 |
| R-squared | 0.177 | -0.309 |

Note: The regression coefficients are the average marginal effects. All regressions control for the covariates including current age, number of primary completed adults, household size, usual monthly household consumption expenditure (in INR), upper caste, religion, month of survey and state fixed effects. The sample is rural children from 12 to 18 years of age. Robust standard errors in parentheses. *** p<0.01, ** p<0.05, * p<0.1.

Table A4: Effect of solid fuel on educational outcomes –Additional sensitivity analysis

| | (1) School attendance-Alternate IV | (2) Years of schooling-Alternate IV | (3) Grade progression-Alternate IV | (4) School attendance-11-18years | (5) Years of schooling-11-18years | (6) Grade progression-11-18years | (7) School attendance-Including kerosene | (8) School attendance-Including kerosene | (9) School attendance-Including kerosene |
|---|---|---|---|---|---|---|---|---|---|
| Solid fuel | -0.211*** | -0.491*** | -0.126*** | -0.186*** | -0.883*** | -0.284*** | | | |
| | (0.052) | (0.114) | (0.036) | (0.056) | (0.129) | (0.043) | | | |
| Dirty fuel | | | | | | | -0.189*** | -1.046*** | -0.322*** |
| | | | | | | | (0.063) | (0.147) | (0.046) |
| Other controls | Y | Y | Y | Y | Y | Y | Y | Y | Y |
| State FE | Y | Y | Y | Y | Y | Y | Y | Y | Y |
| Month of survey FE | Y | Y | Y | Y | Y | Y | Y | Y | Y |
| Observations | 280,339 | 280,011 | 280,011 | 316,097 | 315,761 | 315,761 | 280,339 | 280,011 | 280,011 |
| R-squared | 0.182 | 0.326 | 0.191 | 0.188 | 0.321 | 0.105 | 0.187 | 0.242 | 0.085 |

Note: The regression coefficients are the average marginal effects. All regressions control for the covariates including current age, number of primary completed adults, household size, BPL card, toilet, water in yard, mobile, upper caste, religion, month of survey and state fixed effects. The sample is rural children from 12 to 18 years of age from regression estimates shown in column 1-3 and 7-9. Robust standard errors in parentheses. *** p<0.01, ** p<0.05, * p<0.1.

Figure A5: Time use for fuel collection and preparation and school and homework – Boys

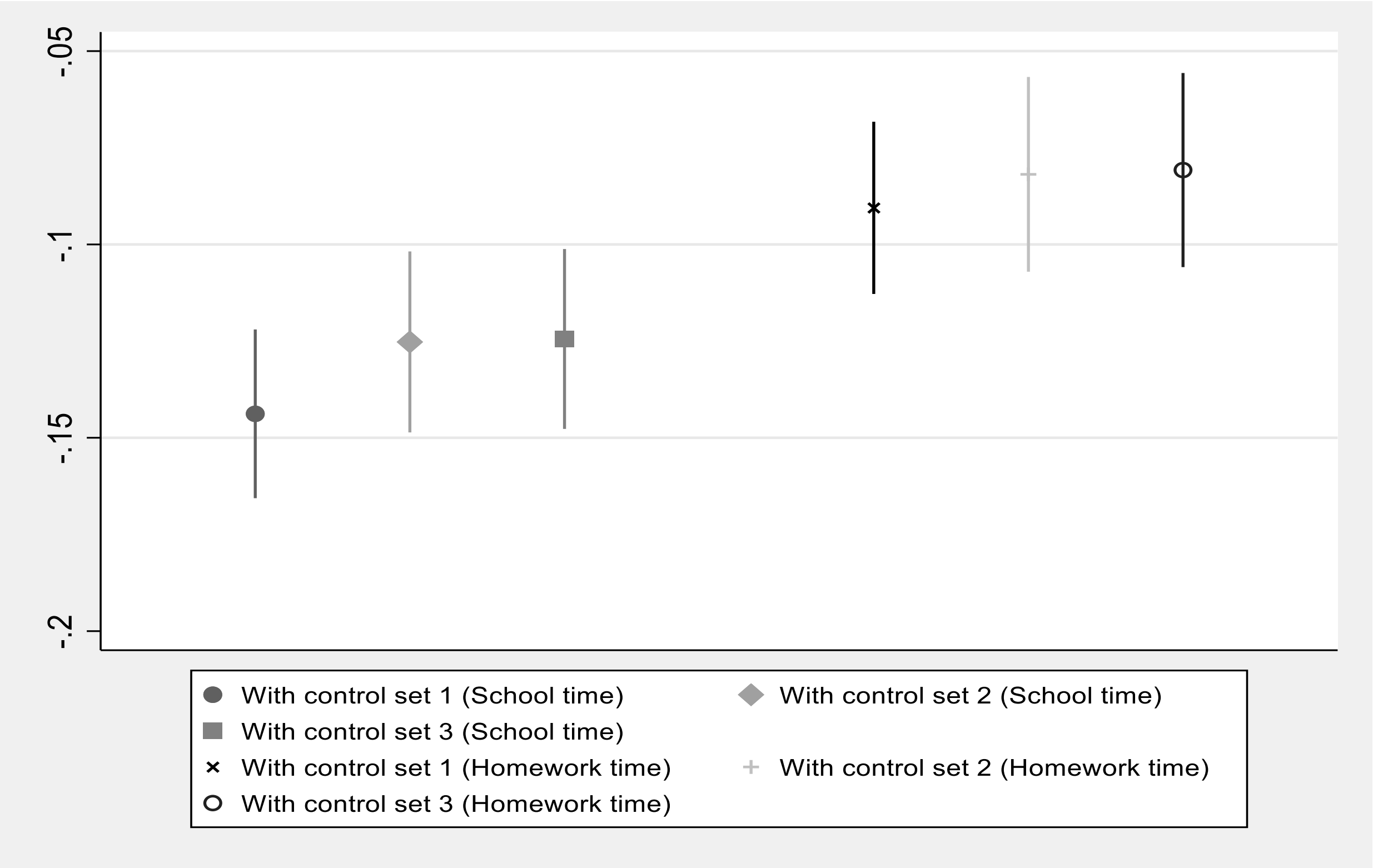

Note: The average marginal effects are plotted along with the 90% confidence intervals. All regressions control for the covariates including current age, household possession of television and toilet, household size, whether the household is cemented or not, whether anyone in the household knows computer usage, whether anyone in the household has a mobile phone, whether the mother attended school or not and state fixed effects. The sample is rural boys from 12 to 16 years of age. Regression output is based on robust standard errors.

Figure A6: Time use for fuel collection and preparation and school and homework – Girls

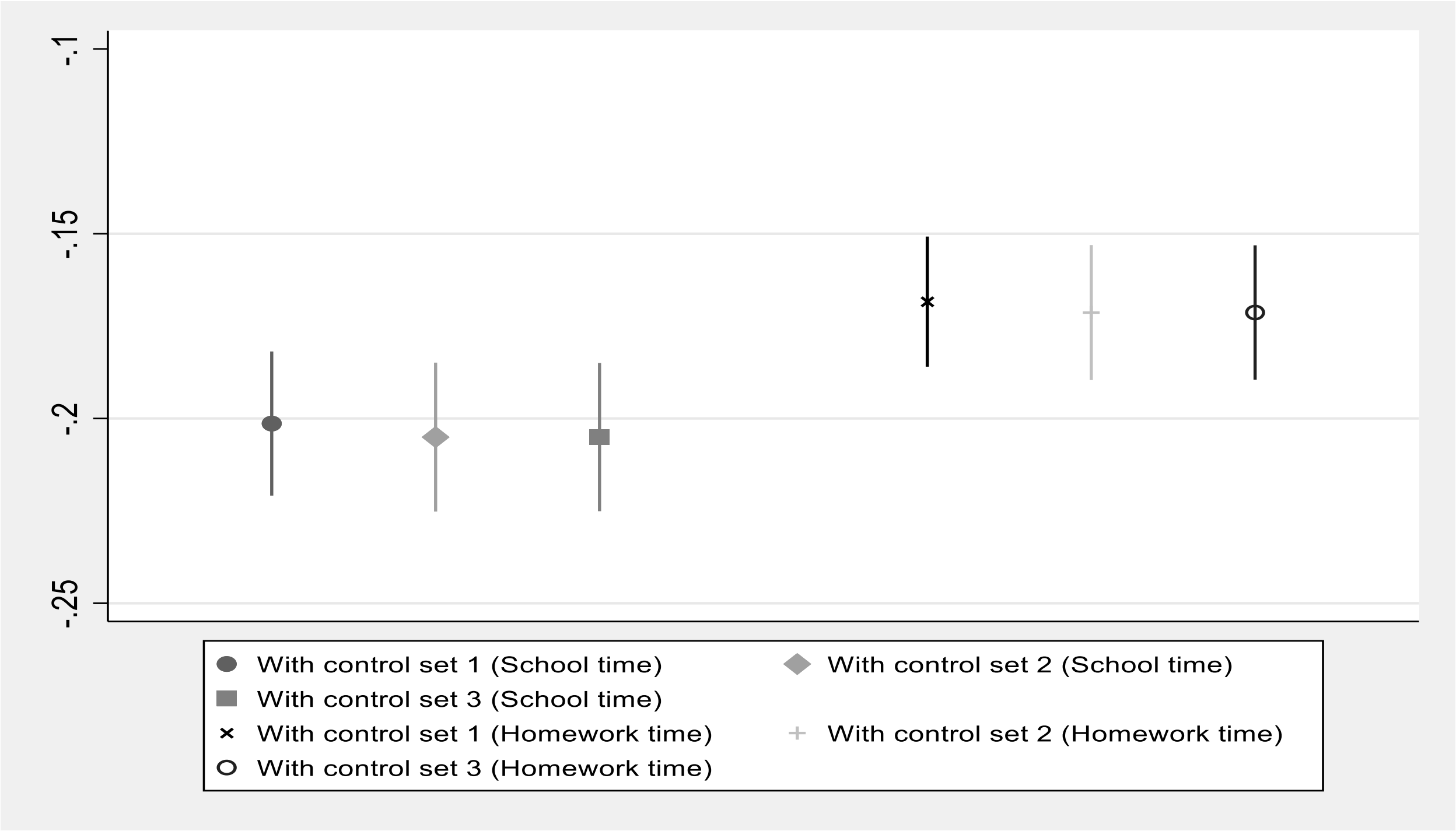

Note: The average marginal effects are plotted along with the 90% confidence intervals. All regressions control for the covariates including current age, household possession of television and toilet, household size, whether the household is cemented or not, whether anyone in the household knows computer usage, whether anyone in the household has a mobile phone, whether the mother attended school or not and state fixed effects. The sample is rural girls from 12 to 16 years of age. Regression output is based on robust standard errors.

Table A7: Health as a possible channel

| | (1)<br>TB[a]- OLS | (2)<br>TB[a]- IV | (3)<br>School attendance-Boys | (4)<br>School attendance-Girls | (5)<br>Years of schooling-Boys | (6)<br>Years of schooling-Girls | (7)<br>Grade progression-Boys | (8)<br>Grade progression-Girls |
|---|---|---|---|---|---|---|---|---|
| Solid fuel | 0.003*** | 0.032 | -0.169** | -0.190** | -0.627*** | -1.427*** | -0.184*** | -0.454*** |
| | (0.001) | (0.026) | (0.081) | (0.091) | (0.182) | (0.228) | (0.056) | (0.073) |
| TB | | | -0.028*** | -0.017** | -0.061*** | -0.048** | -0.021*** | -0.016** |
| | | | (0.008) | (0.008) | (0.018) | (0.021) | (0.006) | (0.007) |
| Other controls | Y | Y | Y | Y | Y | Y | Y | Y |
| State FE | Y | Y | Y | Y | Y | Y | Y | Y |
| Month of survey FE | Y | Y | Y | Y | Y | Y | Y | Y |
| Observations | 184,448 | 184,448 | 143,884 | 136,455 | 143,730 | 136,281 | 143,730 | 136,281 |
| R-squared | 0.008 | 0.002 | 0.164 | 0.212 | 0.337 | 0.133 | 0.152 | 0.003 |

Note: The regression coefficients are the average marginal effects. All regressions control for the covariates including current age, number of primary completed adults, household size, BPL card, toilet, water in yard, mobile, upper caste, religion, month of survey and state fixed effects. [a] Household level regression after controlling for household head's sex and age in addition to above controls but excluding current age. The sample is rural children from 12 to 18 years of age. Robust standard errors in parentheses. *** p<0.01, ** p<0.05, * p<0.1.